\documentstyle[]{article}
\begin{document}
\title{Masses and fields in Microdynamics: a possible foundation for
       dynamic gravity}
\author{W.A. Hofer\\
Dept. of Physics and Astronomy, University College London\\
Gower Street, London WC1E 6BT, UK \quad
Email:w.hofer@ucl.ac.uk}
\date{}

\newcommand{\be}{\begin{eqnarray}\label}
\newcommand{\ee}{\end{eqnarray}}


\maketitle

\begin{abstract}
The quest for a complete theory of microphysics is probably near
the top of the agenda in fundamental physics today. We survey
existing modifications of quantum mechanics to assess their
potential. In the following we present recent results on the
dynamic nature of electric charge. The theoretical model, valid at
all length scales, relates mass oscillations to static electric
and gravity fields. The same concept is used, with substantially
lower frequencies, to compute the intensity of gravity waves
within the solar system. These waves, in the range of kilohertz,
can in principle be detected.
\end{abstract}

\section{In search of new foundations}
A few years ago physicists thought that all the fundamental puzzles of their
field were essentially solved. The experiments in high energy physics,
designed to test the standard model, gave agreement between theoretical
predictions and experimental findings to the n-th decimal, all the
fundamental interactions seemed to be known, the full variety of atomic
behaviour explained; and the whole history of our Universe appeared to
be at the brink of final disclosure.
The first cracks in this monumental edifice of quantum field theory and
general relativity appeared on the level of atoms and, independently, on
the level of galaxies. On the level of atoms it was the remarkable
success of experimental methods in condensed matter physics. These
methods, most prominently the scanning tunnelling microscope, allowed to
study the behaviour of atoms in real time and real space. Henceforth,
chemical processes could not only be mathematically described, they
could actually be observed. Implicitly, this made a reinterpretation
of quantum mechanics necessary. Because if I can study the behaviour
of a {\it single} atom (which is by now routinely done), how can this
experiment be the result of statistics? And if this atom is real,
in what sense precisely is the atom not, which we treat in quantum
mechanics? Exactly the same question can be asked in experiments with
single photons or electrons. Also these experiments are by now routine
in many laboratories around the world. The experimental situation allows
only one answer: quantum mechanics is an algorithm and essentially
incomplete. But then we may immediately ask: Is there a complete theory,
and is there a reality behind the mathematical objects in quantum
mechanics? Since quantum mechanics is the basis for quantum field
theory, these questions are important for the whole of microphysics.
Also, in principle, for high energy physics. So what, we may ask, is
the reality behind a neutrino, a quark, a Higgs boson?

On the level of galaxies the standard model is based on the experimental
fact of an isotropic background radiation in the range of 160 GHz. This
radiation is equal to the radiation of a black body with $T = 2.726$ K
\cite{cobe90}.
It is thought to originate from the birth of our Universe in the ''big bang'',
some 10 to 20 billion years ago. On closer scrutiny this model encountered
several obstacles. The cosmic background radiation, for example, is constant
in all directions. The fact is problematic, because the early universe could
not be completely homogeneous. To reconcile facts with the theory, one either
has to assume a varying velocity of light (in contrast with special relativity),
or a peculiar way, how the universe initially expanded.
Today, the most serious objection against its validity is
motion of galactic mass. According to model calculations the observations
can only be accounted for, if more than 90 percent of a galaxy's mass is
unobservable \cite{rubin97,rubin98}.
But this, in turn, contradicts the calculations of the
upper limit of baryon mass in the big bang model. One either has to resort
to ad hoc hypotheses about the qualities of this ''dark matter'', or
concede, that the big bang model is flawed. But if it is fundamentally
flawed, then the cosmic radiation background must have other origins.
And this, then, could be the point of departure of a completely
different theory about our Universe.

\section{Alternative formulations}

The problems sketched have only surfaced in the last ten to twenty years of
the Twentieth century.
Limiting our discussion to theories amending or extending quantum mechanics,
these fall essentially into two categories: (i) Theories aimed at reconciling
the microscopic with the macroscopic domain. (ii) Theories aimed at recovering
a physical reality behind the mathematical formulations.

It is a constitutional feature of the theories type (i) that they accept the
existing formulations. The aim is therefore not so much an extension of our
knowledge, than limiting the application of existing mathematical expressions.
In this sense, they provide a subset of the unrestricted original frameworks.
Examples of this type are: The consistent histories approach due to Griffiths,
Hartle, and Omnes \cite{griffiths84,hartle91,omnes92}; the
Ghirardi-Rimini-Weber model \cite{ghirardi86}; or the many-world concept
\cite{everett57}. None of these models develops a new physical reality
beyond quantum mechanics, which could in principle be measured.
In this sense, none of the models is suitable to answer the principal
question: What is the physical reality of an atom, photon, electron?

The same deficiency probably afflicts all theories based on classical
mechanics. Because the main addition, in quantum mechanics, is the
uncertainty principle. Mathematically this is expressed by non-commuting
variables. Since this condition inhibits any observation of systems
with a higher degree of precision, it is theoretically necessary to
postulate some hidden variables, which describe fundamental reality.
Mechanics is not developed beyond point-like objects and density
distributions. In particular the notion of a phase is not a
mechanical concept. But as phases play a major role in field physics -
and thus, generally, in microphysics - their omission is a severe,
probably decisive, limitation. This is true before and apart from any analysis of
the actual models used in quantum mechanics. In this sense Bohm's theory
\cite{bohm52}, if it is interpreted as a theory of real particles with
real trajectories, pays the price that also the non-local connections,
inherent in quantum mechanics, must be real. Even though it is therefore
a theory of type (ii), it cannot be consistently formulated as such.

Another theory of type (ii), stochastic electrodynamics \cite{pena96},
focuses on electrodynamics and photons. Classical electrodynamics in
this concept is extended by a zero-point field which carries the
statistical uncertainty inherent in quantum mechanics. Even though
there has recently been new interest in the theory, it is for all
practical purposes restricted to photons. It thus has nothing to say,
so far, about atoms, or electrons.

The most recent theory type (ii), microdynamics
\cite{hofer97,hofer98a,hofer00a,hofer00b}, focuses on wave properties of
matter. These waves are thought to be real. It could be shown that
electrodynamics and quantum mechanics are limiting cases of the
same general model of matter in microphysics. In this theory the
reality behind quantum mechanics is the reality of fields and
densities. Contrary to other theories, microdynamics also allows
to develop a dynamic model of atoms \cite{hofer98b}.

In the initial concept, the model of hydrogen required an ad-hoc
hypothesis: the existence of fields inside the atom, which are
due to density oscillations of the proton. In this sense the
density oscillations were postulated in addition to the electrostatic
charge of the proton. It was only realized recently, that these
oscillations may play a far more fundamental role and in fact
be related to the very existence of electric charge. A line
of research, we shall pursue in the rest of this paper. As
shall be seen shortly, the concept can even be used to bridge
the gap between electrodynamics and gravitation in microphysics.
It thus appears to be at the bottom of what could well become
a completely new field: gravity dynamics.

\section{The nature of charge}

Since the discovery of the electron by J. J. Thomson
\cite{thomson97} the concept of electric charge has remained
nearly unchanged. Apart from Lorentz' extended electron
\cite{lorentz04}, or Abraham's electromagnetic electron
\cite{abraham03}, the charge of an electron remained a point like
entity, in one way or another related to electron mass
\cite{bender84,keller97}. In atomic nuclei we think of charge as a
smeared out region of space, which is structured by the elementary
constituents of nuclear particles, the quarks \cite{montan94}.

But experiments on the quantum hall effect\cite{stormer82,hall90},
performed around 1980, suggested the existence of "fractional
charge" of electrons. Although this effect has later been
explained on the basis of standard theory \cite{laughlin83}, its
implications are worth a more thorough analysis. Because it cannot
be excluded that the same feature, fractional or even continuous
charge, will show up in other experiments, especially since
experimental practice more and more focuses on the properties of
single particles. And in this case the conventional picture, which
is based on discrete and unchangeable charge of particles, may
soon prove too narrow a frame of reference. It seems therefore
justified, at this point, to analyse the very nature of charge
itself. A nature, which would reveal itself as an answer to the
question: What is charge?

With this problem in mind, we reanalyse the fundamental equations
of intrinsic particle properties \cite{hofer98a}. The consequences
of this analysis are developed in two directions. First, we
determine the interface between mechanic and electromagnetic
properties of matter, where we find that only one fundamental
constant describes it: Planck's constant $\hbar$. And second, we
compute the fields of interaction within a hydrogen atom, where we
detect oscillations of the proton density of mass as their source.
Finally, the implications of our results in view of unifying
gravity and quantum theory are discussed and a new model of
gravity waves derived, which is open to experimental tests.

\section{The origin of dynamic charge}

The intrinsic vector field ${\bf E}({\bf r},t)$, the momentum
density ${\bf p}({\bf r},t)$, and the scalar field $\phi ({\bf
r},t)$ of a particle are described by (see \cite{hofer98a}, Eq.
(18)):

\be{1}
 {\bf E}({\bf r},t) = - \nabla \frac{1}{\bar{\sigma}} \,
\phi ({\bf r},t) +  \frac{1}{\bar{\sigma}} \,
\frac{\partial}{\partial t} \, {\bf p}({\bf r},t) \ee

Here $\bar{\sigma}$ is a dimensional constant introduced for
reasons of consistency. Rewriting the equation with the help of
the definitions:

\be{2}
\beta := \frac{1}{\bar{\sigma}} \qquad \beta \phi ({\bf
r},t) := \phi ({\bf r},t) \ee

we obtain the classical equation for the electric field, where in
place of a vector potential ${\bf A}({\bf r},t)$ we have the
momentum density ${\bf p}({\bf r},t)$. This similarity, as already
noticed, bears on the Lorentz gauge as an expression of the energy
principle (\cite{hofer98a} Eqs. (26) - (28)).

\be{3}
{\bf E}({\bf r},t) = - \nabla \, \phi ({\bf r},t) + \beta
\, \frac{\partial}{\partial t} \, {\bf p}({\bf r},t) \ee

Note that $\beta$ describes the interface between dynamic and
electromagnetic properties of the particle. Taking the gradient of
(\ref{3}) and using the continuity equation for ${\bf p}({\bf
r},t)$:

\be{4} \nabla \, {\bf p}({\bf r},t) +  \frac{\partial}{\partial t}
\rho ({\bf r},t) = 0 \ee

where $ \rho ({\bf r},t)$  is the density of mass, we get the
Poisson equation with an additional term. And if we include the
source equation for the electric field ${\bf E}({\bf r},t)$:

\be{5} \nabla \, {\bf E}({\bf r},t) = \sigma ({\bf r},t), \ee

$ \sigma ({\bf r},t) $ being the density of charge, $\epsilon$ set
to 1 for convenience, we end up with the modified Poisson
equation:

\be{6} \Delta \phi ({\bf r},t) = - \underbrace{\sigma ({\bf
r},t)}_{static \, charge} - \underbrace{\beta \,
\frac{\partial^2}{\partial t^2} \rho ({\bf r},t)}_{dynamic \,
charge} \ee

The first term in (\ref{6}) is the classical term in
electrostatics. The second term does not have a classical
analogue, it is an essentially novel source of the scalar field
$\phi$, its novelty is due to the fact, that no dynamic
interpretation of the vector potential $ {\bf A}({\bf r},t)$
exists, whereas, in the current framework, $ {\bf p}({\bf r},t)$
has a dynamic meaning: that of momentum density.

To appreciate the importance of the new term, think of an
aggregation of mass in a state of oscillation. In this case the
second derivative of $\rho$ is a periodic function, which is, by
virtue of Eq. (\ref{6}), equal to periodic charge. Then this
dynamic charge gives rise to a periodic scalar field $\phi$. This
field appears as a field of charge in periodic oscillations: hence
its name, dynamic charge.
We demonstrate the implications of Eq. (\ref{6}) on an easy
example: the radial oscillations of a proton. The treatment is
confined to monopole oscillations, although the results can easily
be generalised to any multipole. Let a proton's radius be a
function of time, so that $r_{p} = r_{p}(t)$ will be:

\be{7} r_{p}(t) = R_{p} + d \cdot \sin \omega_{H} t \ee

Here $R_{p}$ is the original radius, $d$ the oscillation
amplitude, and $\omega_{H}$ its frequency. Then the volume of the
proton $V_{p}$ and, consequently, its density of mass $\rho_{p}$
depend on time. In first order approximation we get:

\be{8} \rho_{p}(t) = \frac{3 M_{p}}{4 \pi} \left(R_{p} + d \sin
\omega_{H} t \right)^{-3} \approx
\rho_{0} \left( 1 - x \sin \omega_{H} t \right) \qquad x :=
\frac{3 d}{R_{p}} \ee

The Poisson equation for the dynamic contribution to proton charge
then reads:

\be{9} \Delta \phi ({\bf r},t) = - \beta x \rho_{0} \omega_{H}^2
\sin \omega_{H} t \ee

Integrating over the volume of the proton we find for the dynamic
charge of the oscillating proton the expression:

\be{10} q_{D}(t) = \int_{V_{p}} d^3 r \beta x \rho_{0}
\omega_{H}^2 \sin \omega_{H} t = \beta x M_{p} \omega_{H}^2 \sin
\omega_{H} t \ee

This charge gives rise to a periodic field within the hydrogen
atom, as already analysed in some detail  and in a slightly
different context \cite{hofer98b}. We shall turn to the
calculation of a hydrogen's fields of interaction in the following
sections. But in order to fully appreciate the meaning of the
dynamic aspect it is necessary to digress at this point and to
turn to the discussion of electromagnetic units.

\section{Natural electromagnetic units}

By virtue of the Poisson equation (\ref{6}) dynamic charge must be
dimensionally equal to static charge, which for a proton is + e.
But since it is, in the current framework, based on dynamic
variables, the choice of $\beta$ also defines the interface
between dynamic and electromagnetic units. From (\ref{10}) we get,
dimensionally:

\be{11} [e] = [\beta] [M_{p} \omega_{H}^2] \quad \Rightarrow \quad
[\beta] = \left[\frac{e}{M_{p} \omega_{H}^2}\right] \ee

The unit of $\beta$ is therefore, in SI units:

\be{12} [\beta] = C \cdot \frac{s^2}{kg} = C \cdot \frac{m^2}{J}
\qquad [SI] \ee

We define now the {\it natural system of electromagnetic units} by
setting $\beta$ equal to 1. Thus:

\be{13} [\beta] := 1 \qquad \Rightarrow \qquad [C] = \frac{J}{m^2}
\ee

The unit of charge C is then energy per unit area of a surface.
Why, it could be asked, should this definition make sense?
Because, would be the answer, it is the only suitable definition,
if electrostatic interactions are accomplished by photons.
Suppose a $\delta^3({\bf r} - {\bf r'})$ like region around ${\bf
r'}$ is the origin of photons interacting with another
$\delta^3({\bf r} - {\bf r''})$ like region around ${\bf r''}$.
Then ${\bf r'}$ is the location of charge. Due to the geometry of
the problem the interaction energy will decrease with the square
of $|{\bf r'} - {\bf r''}|$. What remains constant, and thus
characterises the charge at ${\bf r'}$, is only the interaction
energy per surface unit. Thus the definition, which applies to all
$r^{-2}$ like interactions, also, in principle, to gravity.

Returning to the question of natural units, we find that all the
other electromagnetic units follow straightforward from the
fundamental equations \cite{hofer98a}. However, if we analyse the
units in Lorentz' force equation, we observe, at first glance, an
inconsistency.

\be{22} {\bf F}_{L} = q \left({\bf E} + {\bf u} \times {\bf
B}\right) \ee

The unit on the left, Newton, is not equal to the unit on the
right. As a first step to solve the problem we include the
dielectric constant $\epsilon^{-1}$ in the equation, since this is
the conventional definition of the electric field ${\bf E}$. Then
we have:

\be{23} [{\bf F}_{L}] = \frac{N m}{m^2} \left(\frac{m^4}{N}
\frac{N}{m^3} + \frac{m}{s} \frac{N s}{m^4} \right) = N + N \cdot
\frac{N}{m^4} \ee

Interestingly, now the second term, which describes the magnetic
forces, is wrong in the same manner, the first term was before we
included the dielectric units. It seems thus, that the dimensional
problem can be solved by a constant $\eta$, which is dimensionally
equal to $\epsilon$, and by rewriting the force equation
(\ref{22}) in the following manner:

\be{24} {\bf F}_{L} = \frac{q}{\eta} \left({\bf E} + {\bf u}
\times {\bf B}\right) \qquad [\eta] = N m^{-4} = C m^{-3} =
[\sigma] \ee

The modification of (\ref{22}) has an implicit meaning, which is
worth being emphasise. It is common knowledge in special
relativity, that electric and magnetic fields are only different
aspects of a situation. They are part of a common field tensor
$F_{\mu \nu}$ and transform into each other by Lorentz
transformations. From this point of view the treatment of electric
and magnetic fields in the SI, where we end up with two different
constants ($\epsilon, \mu$), seems to go against the requirement
of simplicity. On the other hand, the approach in quantum field
theory, where one employs in general only a dimensionless constant
at the interface to electrodynamics, the finestructure constant
$\alpha$, is over the mark. Because the information, whether we
deal with the electromagnetic or the mechanic aspect of a
situation, is lost. The natural system, although not completely
free of difficulties, as seen further down, seems a suitable
compromise. Different aspects of the intrinsic properties, and
which are generally electromagnetic, are not distinguished, no
scaling is necessary between ${\bf p}, {\bf E}$ and ${\bf B}$. The
only constant necessary is at the interface to mechanic
properties, which is $\eta$. This also holds for the fields of
radiation, which we can describe by:

\be{26} \phi_{Rad}({\bf r},t) = \frac{1}{8 \pi \eta} \left({\bf
E}^2 + c^2 {\bf B}^2 \right) \ee

Note that in the natural system the usage, or the omission, of
$\eta$ ultimately determines, whether a variable is to be
interpreted as an electromagnetic or a mechanic property. Forces
and energies are mechanic, whereas momentum density is not. The
numerical value of $\eta$ has to be determined by explicit
calculations. This will be done in the next sections.
Comparing with existing systems we note three distinct advantages:
(i) The system reflects the dynamic origin of fields, and it is
based on only three fundamental units: m, kg, s. A separate
definition of the current is therefore obsolete. (ii) There is a
clear cut interface between mechanics (forces, energies), and
electrodynamics (fields of motion). (iii) The system provides a
common framework for macroscopic and microscopic processes.

\section{Interactions in hydrogen}

Returning to proton oscillations let us first restate the main
differences between a free electron and an electron in a hydrogen
atom \cite{hofer98b}: (i) The frequency of the hydrogen system is
constant $\omega_{H}$, as is the frequency of the electron wave.
It is thought to arise from the oscillation properties of a
proton. (ii) Due to this feature the wave equation of momentum
density ${\bf p}({\bf r},t)$ is not homogeneous, but
inhomogeneous:

\be{27} \Delta {\bf p}({\bf r},t) - \frac{1}{u^2}
\frac{\partial^2}{\partial t^2} {\bf p}({\bf r},t) = {\bf f}(t)
\delta^3 ({\bf r}) \ee

for a proton at ${\bf r} = 0$ of the coordinate system. The source
term is related to nuclear oscillations. We do not solve
(\ref{27}) directly, but use the energy principle to simplify the
problem. From a free electron it is known that the total intrinsic
energy density, the sum of a kinetic component $ \phi_{K}$ and a
field component $\phi_{EM}$ is a constant of motion
\cite{hofer98a}:

\be{28} \phi_{K}({\bf r}) + \phi_{EM}({\bf r}) = \rho_{0} u^2 \ee

where $u$ is the velocity of the electron and $\rho_{0}$ its
density amplitude. We adopt this notion of energy conservation
also for the hydrogen electron, we only modify it to account for
the spherical setup:

\be{29} \phi_{K}({\bf r}) + \phi_{EM}({\bf r}) =
\frac{\rho_{0}}{r^2} u^2 \ee

The radial velocity of the electron has discrete levels. Due to
the boundary values problem at the atomic radius, it depends on
the principal quantum number $n$. From the treatment of hydrogen
we recall for $u_{n}$ and $\rho_{0}$ the results \cite{hofer98b}:

\be{30} u_{n} = \frac{\omega_{H} R_{H}}{2 \pi n} \qquad \rho_{0} =
\frac{M_{e}}{2 \pi R_{H}} \ee

where $R_{H}$ is the radius of the hydrogen atom and $M_{e}$ the
mass of an electron. Since $\rho_{0}$ includes the kinetic as well
as the field components of electron ''mass'', e.g. in Eq.
(\ref{29}), we can define a momentum density ${\bf p}_{0}({\bf
r},t)$, which equally includes both components. As the velocity
$u_{n} = u_{n}(t)$ of the electron wave in hydrogen is periodic,
the momentum density ${\bf p}_{0}({\bf r},t)$ is given
by:

\be{31} {\bf u}_{n}(t) = u_{n} \cos \omega_{H} t \, {\bf e}^{r}
\qquad \Longrightarrow \qquad
{\bf p}_{0}({\bf r},t) = \frac{\rho_{0} u_{n}}{r^2} \cos
\omega_{H} t \, {\bf e}^{r} \ee

The combination of kinetic and field components in the variables
has a physical background: it bears on the result that photons
change both components of an electron wave \cite{hofer98a}. With
these definitions we can use the relation between the electric
field and the change of momentum, although now this equation
refers to both components:

\be{33} {\bf E}_{0}({\bf r},t) = \frac{\partial}{\partial t} {\bf
p}_{0}({\bf r},t) = - \frac{\rho_{0} u_{n}}{r^2} \omega_{H} \sin
\omega_{H} t \, {\bf e}^{r} \ee

Note that charge, by definition, is included in the electric field
itself. Integrating the dynamic charge of a proton from Eq.
(\ref{10}) and accounting for flow conservation in our spherical
setup, the field of a proton will be:

\be{34} {\bf E}_{0}({\bf r},t) = \frac{q_{D}}{r^2} = \frac{M_{p}
\omega_{H}^2}{r^2} x \, \sin \omega_{H} t \, {\bf e}^{r} \ee

Apart from a phase factor the two expressions must be equal.
Recalling the values of $u_{n}$ and $\rho_{0}$ from (\ref{30}),
the amplitude $x$ of proton oscillation can be computed. We
obtain:

\be{35} x = \frac{3 d}{R_{p}} = \frac{M_{e}}{(2 \pi)^2 M_{p}}
\cdot \frac{1}{n} \ee

In the highest state of excitation, which for the dynamic model is
$n = 1$, the amplitude is less than $10^{-5}$ times the proton
radius: Oscillations are therefore comparatively small. This
result indicates that the scale of energies within the proton is
much higher than within the electron, say. The result is therefore
well in keeping with existing nuclear models. For higher $n$, and
thus lower excitation energy, the amplitude becomes smaller and
vanishes for $n \rightarrow \, \infty$.

It is helpful to consider the different energy components within
the hydrogen atom at a single state, say $n = 1$, to understand,
how the electron is actually bound to the proton. The energy of
the electron consists of two components.

\be{36} \phi_{K}({\bf r},t) = \frac{\rho_{0} u_{1}^2}{r^2} \sin^2
k_{1} r \cos^2 \omega_{H} t \ee

is the kinetic component of electron energy ($k_{1}$ is now the
wavevector of the wave). As in the free case, the kinetic
component is accompanied by an intrinsic field, which accounts for
the energy principle (i.e. the requirement, that total energy
density at a given point is a constant of motion). Thus:

\be{37} \phi_{EM}({\bf r},t) = \frac{\rho_{0} u_{1}^2}{r^2} \cos^2
k_{1} r \cos^2 \omega_{H} t \ee

is the field component. The two components together make up for
the energy of the electron. Integrating over the volume of the
atom and a single period $\tau$ of the oscillation, we obtain:

\be{38} W_{el} = \frac{1}{\tau} \int_{0}^{\tau} dt \int_{V_{H}}
d^3 r \left(\phi_{K}({\bf r},t) + \phi_{EM}({\bf r},t) \right)
= \frac{1}{2} M_{e} u_{1}^2 \ee

This is the energy of the electron in the hydrogen atom. $W_{el}$
is equal to 13.6 eV. The binding energy of the electron is the
{\it energy difference} between a free electron of velocity
$u_{1}$ and an electron in a hydrogen atom at the same velocity.
Since the energy of the free electron $W_{free}$ is:

\be{39} W_{free} = \hbar \omega_{H} = M_{e} u_{1}^2 \ee

\noindent the energy difference $\triangle W$ or the binding
energy comes to:

\be{40} \triangle W = W_{free} - W_{el} = \frac{1}{2} M_{e}
u_{1}^2 \ee

This value is also equal to 13.6 eV. It is, furthermore, the
energy contained in the photon field $\phi_{Rad}({\bf r},t) $ of
the proton's radiation

\be{41} W_{Rad} = \triangle W = \frac{1}{\tau} \int_{0}^{\tau} dt
\int_{V_{H}} d^3 r \, \phi_{Rad}({\bf r},t) = \frac{1}{2} M_{e}
u_{1}^2 \ee

This energy has to be gained by the electron in order to be freed
from its bond, it is the ionization energy of hydrogen. However,
in the dynamic picture  the electron is not thought to move as a
point particle in the static field of a central proton charge, the
electron is, in this model, a dynamic and oscillating structure,
which emits and absorbs energy constantly via the photon field of
the central proton. In a very limited sense, the picture is still
a statistical one, since the computation of energies involves the
average over  a full period.

\section{The meaning of $\eta$}

The last problem, we have to solve, is the determination of
$\eta$, the coupling constant  between electromagnetic and
mechanic variables. To this end we compute the energy of the
radiation field $W_{Rad}$, using Eqs. (\ref{26}), (\ref{34}), and
(\ref{35}). From (\ref{26}) and (\ref{34}) we obtain:

\be{42} \phi_{Rad}(r,t) = \frac{1}{8 \pi \eta} {\bf E}^2 =
\frac{1}{8 \pi \eta} \cdot \frac{M_{p}^2 \omega_{H}^4}{r^4} x^2
\sin^2 \omega_{H} t \ee

Integrating over one period and the volume of the atom this gives:

\be{43} W_{Rad} = \frac{1}{\tau} \int_{0}^{\tau} dt
\int_{R_{p}}^{R_{H}} 4 \pi r^2 dr \, \phi_{Rad}({\bf r},t)
\approx - \frac{1}{4 \eta} \cdot \frac{M_{p}^2 \omega_{H}^4 x^2}{R_{p}}
\ee

provided $R_{p}$, the radius of the proton is much smaller than
the radius of the atom. With the help of (\ref{35}), and
remembering that $W_{Rad}$ for $n = 1$ equals half the electron's
free energy $\hbar \omega_{H}$, this finally leads to:

\be{44} W_{Rad} = \frac{1}{4 \eta} \cdot \frac{M_{p}^2
\omega_{H}^4 x^2}{R_{p}} = \frac{1}{2} \hbar \omega_{H} \ee

\be{45} \eta = \frac{M_{e}^2 \nu_{H}^3}{2 h R_{p}} = \frac{1.78
\times 10^{20}}{R_{p}} \ee

since the frequency $ \nu_{H}$ of the hydrogen atom equals $6.57
\times 10^{15}$ Hz. Then $\eta$ can be calculated in terms of the
proton radius $R_{p}$. This radius has to be inferred from
experimental data, the currently most likely parametrisation being
\cite{eisberg85}:

\be{46} \frac{\rho_{p}(r)}{\rho_{p,0}} = \frac{1}{1 + e^{(r -
1.07)/0.55}} \ee

radii in fm. If the radius of a proton is defined as the radius,
where the density $\rho_{p,0}$ has decreased to $\rho_{p,0}/e$,
with e the Euler number, then the value is between 1.3 and 1.4 fm.
Computing $4 \pi$ the inverse of $\eta$, we get, numerically:

\begin{eqnarray} \frac{4 \pi}{\eta} &=& 0.92 \times 10^{-34} \qquad (R_{p} =
1.3 fm) \nonumber \\ &=& 0.99 \times 10^{-34} \qquad (R_{p} = 1.4
fm)
\\ & = & 1.06 \times 10^{- 34} \qquad (R_{p} = 1.5 fm) \nonumber \ee

Numerically, this value is equal to the numerical value of
Planck's constant $\hbar$ \cite{uip78}:

\be{47} \hbar_{UIP} = 1.0546 \times 10^{-34} \ee

Given the conceptual difference in computing the radius the
agreement seems remarkable. Note that this is a genuine derivation
of $\hbar$, because nuclear forces and radii fall completely
outside the scope of the theory in its present form. If
measurements of $R_{p}$ were any different, then we would be
faced, at this point, with a meaningless numerical value.
Reversing the argument it can be said, that the correct value - or
rather the meaningful value - is a strong argument for the
correctness of our theoretical assumptions. For the following,
we redefine the symbol $\hbar$:

\be{54} \hbar := 1.0546 \times 10^{-34} [N^{-1} m^4] \ee

Then we can rewrite the equations for ${\bf F}$, the Lorentz
force, for ${\bf L}$, angular momentum related to this force, and
$\phi_{Rad}$, the radiation energy density of a photon in a very
suggestive form:

\be{55} {\bf F} = \hbar q \left(\frac{\bf E}{4 \pi} + {\bf u}
\times \frac{\bf B}{4 \pi} \right) \qquad
        {\bf L} = \hbar q \,{\bf r} \times \left(\frac{\bf E}{4
\pi} + {\bf u} \times \frac{\bf B}{4 \pi} \right) \ee

\be{57} \phi_{Rad} = \frac{\hbar}{2} \left[\left(\frac{\bf E}{4
\pi}\right)^2 + c^2 \left(\frac{\bf B}{4 \pi}\right)^2 \right] \ee

Every calculation of mechanic properties involves a multiplication
by $\hbar$. Since $\hbar$ is a scaling constant, the term
''quantization'', commonly used in this context, is misleading.
Furthermore, it is completely irrelevant, whether we compute an
integral property (the force in (\ref{55})), or a density
($\phi_{Rad}$ in (\ref{57}), a force density can also be obtained
by replacing charge q by a density value). What is, in a sense,
discontinuous, is only the mass contained in the shell of the atom. But
this mass depends, as does the amplitude of $\phi_{Rad}(r,t)$, on
the mass of the atomic nucleus. Thus the only discontinuity left on
the fundamental level, is the mass of atomic nuclei. That the
energy spectrum of atoms is discrete, is a trivial observation in
view of boundary conditions and finite radii.

All our calculations so far focus on single atoms. To get the
values of mechanic variables in SI units used in macrophysics, we
have to include the scaling between the atomic domain and the
domain of everyday measurements. Without proof, we assume this
value to be $N_{A}$, Avogadro's number. The scale can be made
plausible from solid state physics, where statistics on the
properties of single electrons generally involve a number of
$N_{A}$ particles in a volume of unit dimensions
\cite{ashcroft74}. And a dimensionless constant does not show up
in any dimensional analysis.

\section{Dynamic gravity fields}

The dynamic fields of mass in motion, and the interpretation of
charge as a dynamic feature of mass has far-reaching consequences.
In general relativity the connection between electrodynamics
and gravity is obtained only via the energy stress tensor $T_{ab}$,
according to Einstein's equation \cite{wald84}:

\begin{equation}
G_{ab} = 8 \pi T_{ab}
\end{equation}

Here $T_{ab}$ is related in a complicated manner to the electromagnetic
fields at a given position. $G_{ab}$ describes the curvature
of spacetime at the same position, which in turn is related to gravity.
There is no direct way, in Einstein's theory, from the electromagnetic
to the gravitational properties of matter. Indirect routes have been
explored in the past \cite{wald72,hawking75,wald75,sidharth98},
focusing on the notion of black holes and a length scale considerably
below the atomic domain. To date, the only connection between microphysics
and gravity is thought to exist in extreme environments, like the first
few seconds after the big bang or the vicinity of a black hole.
The interpretation of electric charge as a dynamic feature of mass,
elaborated in the previous sections, allows a connection in standard
situations. For the first time we may ask, whether the gravitational
aspect of matter (= its mass) and the electromagnetic aspect
(= its charge) are only different scales of the same physical
feature: its oscillations.

Oscillations couple strongly to the environment of matter, therefore
transport, dissipation and radiation effects should in principle
alter the picture in a dynamic way. The fundamental question, in
any such theory, is exactly the opposite one posed in static concepts.
Static theories have difficulties with the question: How do things change?
Dynamic theories, on the other hand, must answer the following one:
How can things remain stable?

An answer to this puzzle could lie in the frequency, thus the time scale
involved. It is well known that biological organisms utilise mainly
electromagnetic or chemical interactions, the time scale for the life
of these organisms is in the range of days to years. The frequencies
involved in the interactions are around $10^{14}$ to $10^{15}$ Hz.
The frequency is also the main parameter for the amplitude of
the field of dynamic charge. From Eq. (\ref{34}) we get:

\be{59} |{\bf E}| = \frac{q_{D}}{r^2} \approx
\frac{M \omega^2}{r^2} \ee

The coupling between mass and gravity is much weaker than between
charge and its electrostatic field. We have two ways, in principle,
to estimate the difference: either we set mass and charge equal to
1 and compute the fields. Then the ratio between gravity and
electrostatics is $\epsilon_{0} G \approx 10^{-22}$. For the frequency
of the dynamic gravity field we get consequently:

\be{60} \omega_{G} \approx 10^{-11} \omega_{E} \ee

Or we compute the ratio of the fields from the static equations,
accounting for physical units by a constant $k_{u}$, which shall
be purely dimensional. Then we have:

\be{61} |{\bf E}| = \frac{e}{4 \pi \epsilon_{0} r^2}
        := M \frac{\omega_{E}^2}{r^2} \qquad
        |{\bf G}| = G \frac{M}{r^2}
        := M \frac{\omega_{G}^2}{r^2}   \qquad
        \omega_{G} = \sqrt{k_{u} \frac{4 \pi \epsilon_{0} G M}{e} }
                     \omega_{E} \ee

At present, we have no way to remove this ambiguity about
the exact numerical value. Consequently, the scale $\eta$ between
the two frequencies will be:

\be{64} \omega_{G} = \eta \cdot \omega_{E} \qquad
        10^{-14} \le \eta \le 10^{-11} \ee

Here $\omega_{E}$ is the characteristic electromagnetic frequency,
$\omega_{G}$ its gravitational counterpart. The hypothesis can in
principle be tested. If $\nu_{G}$, the frequency of gravity radiation,
is about $\eta$ times the frequency of proton oscillation, we get:

\be{65} \nu_{G} \approx 10 \mbox{Hz} - 10 \mbox{kHz} \ee

The frequency is in the same range as the theoretical results based
on general relativity \cite{gravwave00}. Also the implications are
the same: if fields of this frequency range exist in space, we would
attribute these fields to stellar gravity. However, since there is no
difference between electromagnetic waves and gravitational waves,
apart from their frequency, we would also attribute electromagnetic
waves in this range to gravity. To estimate the intensity of this,
hypothetical, field, we use Eq. (\ref{33}):

\be{66} {\bf G}_{S}({\bf r},t) = \frac{\partial}{\partial t} {\bf
p}_{E}({\bf r},t) \ee

Here ${\bf G}_{S}$ is the solar gravity field. The momentum
density and its derivative can be inferred from centrifugal
acceleration.

\be{67} \frac{\partial}{\partial t} \, {\bf p}_{E}({\bf r},t) =
\rho_{E} \, a_{C} \, {\bf e}^{r} \qquad \rho_{E} = \frac{3
M_{E}}{4 \pi R_{E}^3} \qquad a_{C} = \omega_{E}^2 \, R_{O} \ee

where $R_{O}$ is the earth's orbital radius and where we have
assumed isotropic distribution of terrestrial mass. Then Eq.
(\ref{42}) leads to:

\be{68} \phi_{G}(r = R_{O}) = \frac{\hbar}{2} \left(
\frac{G_{S}}{4 \pi} \right)^2 = \frac{\hbar}{2} \left( \frac{3
M_{E} R_{O}}{4 R_{E}^3 \tau_{E}^2} \right)^2 \ee

Note the occurrence of Planck's constant also in this equation,
although all masses and distances are astronomical. The intensity
of the field, if calculated from (\ref{68}), is very small. To
give it in common measures, we compute the flow of gravitational
energy through a surface element at the earth's position. In SI
units we get:

\be{69} J_{G} (R_{O}) = \phi_{G}(R_{O}) \cdot N_{A} \cdot c
\approx 70 mW/m^2 \ee

Compared to radiation in the near visible range - the solar
radiation amounts to over 300 Watt/m$^2$ \cite{morrison96} - the
value seems rather small. But considering, that also radiation in
the visible range could have an impact on terrestrial motion, the
intensity of the gravity waves could be, in fact, much higher.
Concerning the detectors of gravity waves, by now in operation at
several locations around the world \cite{gravwave00},
the hypothesis involves
a conjecture: not, that gravity waves have not been detected
so far, because their intensity is so small, but because they are
so ubiquitous. The background noise, all experimental groups
report as a major obstacle, could be the main experimental feature
of gravity waves. If this is the case, then an unambiguous detection
is not possible on earth. To that end, electromagnetic radiation in
space is the only possible source for analysis.

Coming back to the question, how the solar system in a dynamic
theory of gravity could be stable, we find that the frequency
scale between gravity and electricity is very different. If the
typical lifetime of a primitive organism based on chemical
interactions is in the range of days, then the typical lifetime
of an organism based on gravity would be in the range of
billion years. This value is in the same order of magnitude
as current estimates in cosmology about the typical lifetime
of a small yellow star like our sun.

\section*{Acknowledgements}

Thanks are due to Dr. Sidharth and the B M Birla Foundation for the invitation.
Financial support by the University College London is gratefully acknowledged.



\end{document}